\documentclass[preprint,preprintnumbers,amsmath,amssymb,showkeys]{revtex4-1}
\usepackage[dvips]{graphicx}
\usepackage{latexsym}
\usepackage{natbib}
\usepackage[usenames]{color}
\usepackage{multirow}
\usepackage{dsfont}

\definecolor{Blue}{rgb}{0,0,0.6}
\definecolor{Green}{rgb}{0,0.6,0}
\definecolor{Red}{rgb}{0.6,0,0}

\begin{document}

\title{Backtracking activation impacts the criticality of excitable networks}

\author{Renquan Zhang}
\affiliation{School of Mathematical Sciences, Dalian University of Technology, Dalian 116024, China}

\author{Guoyi Quan}
\affiliation{School of Mathematical Sciences, Dalian University of Technology, Dalian 116024, China}

\author{Jiannan Wang}
\email{wangjiannan@buaa.edu.cn}
\affiliation{Research Institute of Frontier Science, Beihang University, Beijing 100191, China\\
Key Laboratory of Mathematics Informatics Behavioral Semantics, Ministry of Education, China}

\author{Sen Pei}
\email{sp3449@cumc.columbia.edu}
\affiliation{Department of Environmental Health Sciences, Mailman School of Public Health, Columbia University, New York, NY 10032, USA}

\begin{abstract}
Networks of excitable elements are widely used to model real-world biological and social systems. The dynamic range of an excitable network quantifies the range of stimulus intensities that can be robustly distinguished by the network response, and is maximized at the critical state. In this study, we examine the impacts of backtracking activation on system criticality in excitable networks consisting of both excitatory and inhibitory units. We find that, for dynamics with refractory states that prohibit backtracking activation, the critical state occurs when the largest eigenvalue of the weighted non-backtracking (WNB) matrix for excitatory units, $\lambda^E_{NB}$, is close to one, regardless of the strength of inhibition. In contrast, for dynamics without refractory state in which backtracking activation is allowed, the strength of inhibition affects the critical condition through suppression of backtracking activation. As inhibitory strength increases, backtracking activation is gradually suppressed. Accordingly, the system shifts continuously along a continuum between two extreme regimes -- from one where the criticality is determined by the largest eigenvalue of the weighted adjacency matrix for excitatory units, $\lambda^E_W$, to the other where the critical state is reached when $\lambda_{NB}^E$ is close to one. For systems in between, we find that $\lambda^E_{NB}<1$ and $\lambda^E_W>1$ at the critical state. These findings, confirmed by numerical simulations using both random and synthetic neural networks, indicate that backtracking activation impacts the criticality of excitable networks.
\end{abstract}


\maketitle

\section{Introduction}

Excitable networks have been used to model a range of phenomena in biological and social systems including signal propagation in neural networks~\cite{kinouchi2006optimal,copelli2005signal,gollo2009active,gollo2016diversity,wang2017approximate,kinouchi2019stochastic}, information processing in brain networks~\cite{gautam2015maximizing,williams2014quasicritical,marro2013signal}, epidemic spread in human population and information diffusion in social networks~\cite{karrer2011competing,van2012epidemic,dodds2013limited,pei2015detecting}. The collective dynamics of excitable nodes enable the networked system to distinguish stimulus intensities varied by several orders of magnitude, characterized by a large dynamic range in response to external stimuli. In previous studies, it was found that, for a number of excitable network models, the dynamic range is maximized at the critical state~\cite{kinouchi2006optimal,larremore2011predicting,larremore2011effects,copelli2005intensity,zhang2018dynamic}. As a result, understanding the condition of criticality is essential for improving the performance and functionality of excitable networks.

The critical condition for excitable networks composed of only excitatory nodes has been extensively studied. In homogeneous random networks, the critical state corresponds to a unit branching ratio~\cite{kinouchi2006optimal}. For more general network structures, the criticality for dynamics without refractory state is characterized by the unit largest eigenvalue of the weighted adjacency matrix~\cite{larremore2011predicting}. More recently, it was shown that for dynamics with refractory states, the critical state is governed by the largest eigenvalue of the weighted non-backtracking (WNB) matrix~\cite{zhang2018dynamic}. In these studies, the largest eigenvalue of the weighted adjacency matrix or WNB matrix is used to define the critical state of excitable networks. However, when inhibitory nodes are introduced, it is unclear how criticality is related to the largest eigenvalues of these two matrices.

In this study, we explore the critical condition of excitable networks consisting of both excitatory (E) and inhibitory (I) nodes. Inhibition presents in many real-world systems and plays a critical role in model dynamics and functions~\cite{adini1997excitatory,park2010irregular,folias2011new,pei2012how,larremore2014inhibition,mongilli2018inhibitory}. For instance, the introduction of inhibitory nodes into an excitable network operating near the critical state leads to self-sustained network activity~\cite{larremore2014inhibition}, and inhibitory connectivity may be essential in maintaining long-term information storage in volatile cortex~\cite{mongilli2018inhibitory}. In order to elucidate the relationship between criticality and the largest eigenvalues of the weighted adjacency matrix and WNB matrix, we study an excitatory-inhibitory (EI) network model equipped with a threshold-like activation rule~\cite{larremore2014inhibition}. Specifically, we focus on the impact of backtracking activation paths on the critical condition.

We first analyze the model dynamics of EI networks in two extreme conditions where backtracking activation is allowed without restrictions or entirely prohibited. We find that, in the former case, the critical state is better characterized by the largest eigenvalue of the weighted adjacency matrix for excitatory nodes, $\lambda^{E}_{W}$, while in the latter case, the criticality is more related to the largest eigenvalue of the WNB matrix for excitatory nodes, $\lambda^{E}_{NB}$. For EI models with refractory states that preclude backtracking activation, the critical state is achieved when $\lambda^{E}_{NB}$ is close to one. For EI models without refractory state (i.e., with only resting and excited states), however, the analytical form of the critical condition becomes intractable. We show that, qualitatively, the system gradually shifts from the former case to the latter case as the strength of inhibition increases: for negligible inhibition, $\lambda^{E}_{W}$ is closer to one at the critical state; for strong inhibition, $\lambda^{E}_{NB}$ is closer to one at the critical state; for moderate inhibition, we find $\lambda^E_{NB}<1$ and $\lambda^E_W>1$ at the critical state. Using numerical simulations in both random and synthetic networks, we verify that a larger inhibitory strength tends to suppress more backtracking activation, which explains the transition between these two regimes. Our findings highlight the impact of backtracking activation, a form of dynamical resonance, on the criticality of excitable networks, and may provide new insight into the study of similar dynamical processes in networked systems.

\section{Materials and methods}

\subsection{The excitable network model}

We consider excitable networks consisting of both excitatory (E) and inhibitory (I) nodes~\cite{larremore2014inhibition}. Contrary to the function of excitatory nodes, the effect of inhibitory nodes is to decrease the activation probability of their neighbors once they are activated. In a network with $N$ nodes, we use $s_i(t)$ to represent the state of node $i$ at time $t$. Both types of nodes can be in one of $m+1$ states: the resting state $s_i(t)=0$, the excited state $s_i(t)=1$, and refractory states $s_i(t)=2, 3,\cdots,m$. At each discrete time $t$, a resting node can be activated by an external stimulus with a probability $\eta$, or activation propagation from its neighbors independently. Specifically, the signal input strength from a node $j$ to a neighboring node $i$, denoted by $a_{ij}$, satisfies $a_{ij}>0$ if node $j$ is excitatory and $a_{ij}<0$ if node $j$ is inhibitory. If node $j$ and node $i$ are not connected in the network, we set $a_{ij}=0$. The weighted adjacency matrix $A=\{a_{ij}\}_{N\times N}$ thus fully describes the network structure as well as the signal input strength between all pairs of nodes. If a resting node $i$ is not activated by the external stimulus, its activation probability in the next time step $t+1$ is calculated by summing inputs from all neighbors through a transfer function $\sigma(\cdot)$:
\begin{equation}
s_i(t+1)=1 \text{ with probability } \sigma\left(\sum_{j=1}^{N}a_{ij}\tau(s_j(t))\right).
\end{equation}
Here $\sigma(\cdot)$ is a piecewise linear function: $\sigma(x)=0$ for $x\leq0$; $\sigma(x)=x$ for $0<x<1$; and $\sigma(x)=1$ for $x\geq1$. $\tau(\cdot)$ is a characteristic function: when $s_j(t)=1$, $\tau(s_j(t))=1$; otherwise, $\tau(s_j(t))=0$. A node can be activated by its neighbors if the net input is positive. Once activated, node $i$ will transit to refractory states deterministically. That is, $s_i(t+1)=s_i(t)+1$ if $1\leq s_i(t)<m$ and $s_i(t+1)=0$ if $s_i(t)=m$. Note that, if $m=1$, there will be no refractory state and the node will directly return to the resting state after activation.

In this study, we use {\it undirected} networks in which signals can be transmitted in both directions, and assume that the number of E nodes $N_e$ is larger than the number of I nodes $N_i$. Empirical studies on real-world excitable systems reveals that the majority of neural inputs are excitatory~\cite{soriano2008development}. For ease of analysis, we rearrange node indices so that nodes with index $1\leq i\leq N_e$ are excitatory and the rest are inhibitory. We consider both homogeneous and heterogeneous networks. For homogeneous network structure, we first generate two Erd\H{o}s-R\'{e}nyi (ER) random networks consisting of $N_e$ excitatory nodes and $N_i$ inhibitory nodes. Within each network, each pair of nodes is connected with a probability $\alpha$. We then randomly connect E nodes and I nodes with a probability $\beta$. For heterogeneous network structure, two scale-free (SF) networks of E nodes and I nodes with a power-law degree distribution $P(k)\propto k^{-\gamma}$ are generated using the configuration model~\cite{molloy1995critical}. These two networks are then connected by randomly linking E nodes and I nodes with a probability $\beta$. In other words, networks of same types of nodes are constructed using either an ER model with a pairwise linking probability $\alpha$ or a configuration model for SF networks with a power-law exponent $\gamma$. Different types of nodes are connected randomly using a pairwise linking probability $\beta$. We assume the absolute values of link weights $|a_{ij}|$ are distributed uniformly within a range, and the effect of inhibitory nodes is solely represented by the connections from I to E nodes.

In real-world excitable systems, the majority of units are excitatory. For instance, in cortex, approximately 20\% of neuron inputs are inhibitory~\cite{soriano2008development}. Together with the fact that the inhibitory nodes need to be activated first before their release of inhibitory signals, the effect of I-I interactions is quite nominal in response to weak external stimuli. In some modeling studies, the I-I interactions were even neglected~\cite{stefanescu2008low}. In addition, we focus on the relative strength of same-type and cross-type interactions. It would be sufficient to keep one constant and vary the other (i.e., $\beta$). Considering these factors, we decided to study the effect of a varying $\beta$.

\subsection{Dynamic range and criticality}

\begin{figure}
\centering
\includegraphics[width=0.8\columnwidth]{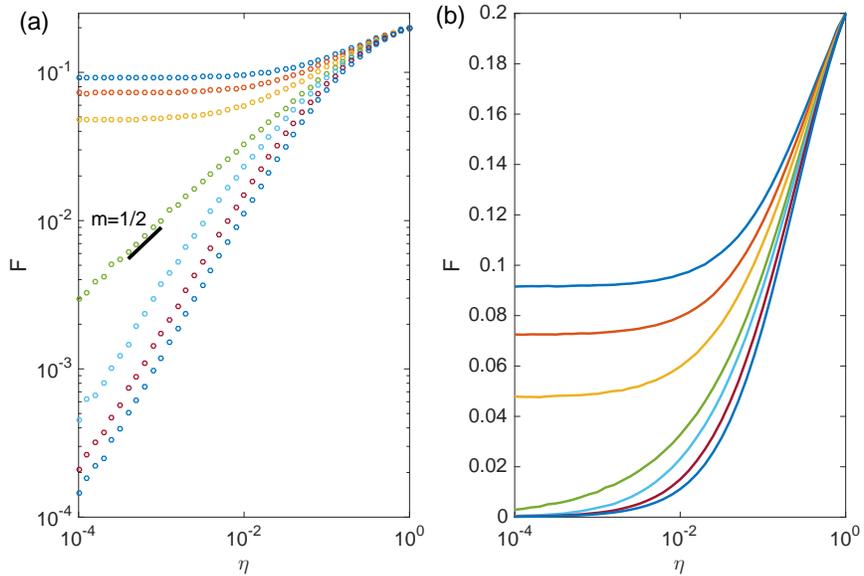}\\
\caption{Network response $F$ in response to external stimulus intensity $\eta$. We show the response curves for an EI network constructed by connecting two ER networks ($N_e=3,000$, $N_i=2,000$, $\alpha=3\times 10^{-3}$, $\beta=1\times 10^{-3}$, $m=4$). The absolute link weights $|a_{ij}|$ are drawn uniformly from 0.1 to 0.2. We multiply link wights with different values to adjust the system to stay in the subcritical, critical and supercritical states. The network response $F$ is shown in logarithmic scale in (a) and linear scale in (b). At the critical state, we show that $F(\eta)\propto\eta^{1/2}$ in (a). The dynamic range of the system is calculated based on the response curve.}\label{F_eta}
\end{figure}

The dynamic range of an excitable network measures the range of stimulus intensities that are distinguishable based on network response~\cite{kinouchi2006optimal}. For a given stimulus intensity $\eta\in(0,1)$, the network response $F$ is defined as
\begin{equation}
F=\lim_{T\to\infty}\frac{1}{T}\sum_{t=0}^TS^t,
\end{equation}
where $S^t$ is the fraction of excited nodes at time $t$. The response $F(\eta)$ increases monotonically with a growing intensity of external stimulus $\eta$ in a nonlinear fashion (see figure~\ref{F_eta}). For a strong stimulus intensity $\eta\to1$, the response $F(\eta)$ will saturate and retain at a maximum value $F_{max}=1/(m+1)$. For a negligible stimulus intensity $\eta\to0$, the minimum response $F_0$ depends on the state of the excitable system. In subcritical state, $F(\eta)$ is a linear function of $\eta$ for $\eta\to0$, i.e., $F(\eta)\propto\eta$, with $F_0\to0$. At the critical state, $F_0$ still approaches to zero but the function $F(\eta)$ becomes nonlinear: $F(\eta)\propto\eta^{1/2}$ (see figure~\ref{F_eta}(a)). The exponent 1/2 is called the Steven's exponent, which characterizes the criticality of the collective dynamics~\cite{kinouchi2006optimal,copelli2002physics}. In supercritical state, the excitation caused by external stimulus can be self-sustained. Therefore, $F_0$ becomes a positive number.

The dynamic range of an excitable network is defined based on the function $F(\eta)$. In particular, we define dynamic range $\Delta$ as
\begin{equation}
\Delta=10\log_{10}\frac{\eta_{high}}{\eta_{low}},
\end{equation}
where $\eta_{high}$ and $\eta_{low}$ are stimulus intensities corresponding to network responses $F_{high}$ and $F_{low}$ (here $F_{x}=F_0+x(F_{max}-F_0)$ for $x\in[0,1]$). In this study, we use $\eta_{0.9}$ and $\eta_{0.1}$, discarding stimuli that are too close to saturation or too weak to be distinguished from $F_0$. Previous studies have demonstrated that dynamic range is maximized at the critical state of an excitable system~\cite{kinouchi2006optimal,larremore2011predicting}. Without forcing of external stimuli, excitation activity will eventually die out in subcritical state but become self-sustained above the critical point. This feature allows us to identify the critical point using the maximization of dynamic range.

The criticality of the proposed model is closely related to percolation. As pointed out in a previous study~\cite{newman2002spread}, activation or disease transmission in networks can be mapped to a bond percolation process. In models with only excitatory units, an absorbing state of activation extinction (or a negligible giant component in percolation) exists for a low transmission probability. The criticality of the system corresponds to the transition point of the stability of this absorbing state, which is determined by the branching ratio~\cite{kinouchi2006optimal}, defined as the average number of secondary activations induced by one excited node. In disease transmission, this quantity is referred to as the reproductive number. If the branching ratio is below one, the absorbing state is stable as all activations eventually die out; in contrast, if the branching ratio is larger than one, the absorbing state becomes unstable and any initial activation will amplify and converge to a non-zero absorbing state. This stability change is equivalent to the emergence of a non-zero giant component in percolation. The introduction of inhibitory units reduces the transmission probability of activation, thus delays the emergence of a non-zero absorbing state. In particular, inhibitory inputs can be viewed as a means of regulation of the system, and how inhibition is imposed on excitatory units could affect when the stability transition occurs. The phase transition in systems with both excitatory and inhibitory units therefore depends on the competition and interplay between excitatory and inhibitory signals. Depending on whether an excited node can be activated immediately after its last excitation, the effect of inhibitory units is different. This study aims to explore how inhibitory units would change the condition of criticality.

\section{Results}

The number of refractory states in model dynamics determines whether backtracking activation is permitted. Backtracking activation describes the following phenomenon of dynamical resonance: an excitatory node $i$ activated at time $t$ increases the activation probability of its excitatory neighbors at time $t+1$, which in turn increases the activation probability of node $i$ at time $t+2$. This behavior is only possible when there is no refractory state (i.e., $m=1$) so that excited nodes can directly return to the resting state at time $t+1$. For dynamics with refractory states (i.e., $m>1$), nodes excited at time $t$ will enter refractory states at time $t+1$ thus cannot be activated again at time $t+2$. Following this dynamical rule, any backtracking activation is prohibited.

\subsection{Dynamics without backtracking activation}

We first analyze the simpler case where backtracking is precluded by the existence of refractory states (i.e., $m>1$). To account for the dynamics without backtracking, we formulate the model evolution in a message-passing framework~\cite{zhang2018dynamic}, which is frequently used in statistical physics and network science~\cite{mezard2009information,morone2015influence,pei2017efficient}. For a link from $i$ to $j$ ($i\to j$), we create a ``cavity'' at node $j$ by ``virtually'' removing it from the network, and examine the probability of node $i$ being activated in the absence of node $j$, denoted by $p_{i\to j}^t$ at time $t$. The procedure of creating a virtual cavity at node $j$ blocks the backtracking path $i\to j\to i$, and therefore excludes the contribution via the consecutive activation $i\to j\to i$ to the activation probability of node $i$. This framework precisely depicts the model dynamics with refractory states.

For sparse networks without too many short loops, the probabilities $p_{i\to j}^t$ for neighboring nodes are mutual independent. Under this condition, the probability $p_{i\to j}^{t}$ for each node $i$ can be recursively written as follows:
\begin{equation}\label{activation_nobacktrack}
p_{i\to j}^{t+1}=(1-\sum_{l=0}^{m-1}p_{i\to j}^{t-l})\left[\eta+(1-\eta)\sigma\left(\sum_{k\in\partial i\setminus j}a_{ik}p_{k\to i}^t\right)\right].
\end{equation}
Here $\partial i\setminus j$ is the set of neighbors of node $i$ excluding $j$. The probability that node $i$ is excited at time $t+1$, denoted by $p_i^{t+1}$, is calculated by putting node $j$ back to the network:
\begin{equation}\label{activation_nobacktrack1}
p_{i}^{t+1}=(1-\sum_{l=0}^{m-1}p_{i}^{t-l})\left[\eta+(1-\eta)\sigma\left(\sum_{k\in\partial i}a_{ik}p_{k\to i}^t\right)\right].
\end{equation}
Note that, although Eqs~(\ref{activation_nobacktrack})-(\ref{activation_nobacktrack1}) are derived for locally tree-like sparse networks, it has been found that results based on the sparseness assumption work well even for some networks with dense clusters~\cite{melnik2011unreasonable}.

\subsubsection{Analysis in the case of negligible inhibition}

The piecewise transfer function $\sigma(\cdot)$ imposes a threshold-like activation rule that depends on the collective dynamics of all neighbors. Because the value of net input $\sum_{k\in\partial i\setminus j}a_{ik}p_{k\to i}^t$ is unknown, it becomes complicated to expand the right-hand-side of Eq~(\ref{activation_nobacktrack}) except for some extreme cases. Here, we consider a special case where the cross-type interaction $\beta$ is negligible, i.e., $\beta\to 0$. Under this extreme condition, excitatory and inhibitory nodes in effect form two nearly separate communities. In particular, inhibitory nodes are unlikely to be activated in response to weak stimuli as they almost only receive signals from inhibitory peers. As a consequence, it is suffice to consider only excitatory nodes to compute network response.

In the steady state, denote the limiting probabilities as $\lim_{t\to\infty}p^t_{i\to j}=p_{i\to j}$ and $\lim_{t\to\infty}p^t_i=p_i$. For excitatory nodes, Eqs~(\ref{activation_nobacktrack})-(\ref{activation_nobacktrack1}) becomes
\begin{equation}\label{activation_nobacktrackE}
p_{i\to j}=(1-mp_{i\to j})\left[\eta+(1-\eta)\sum_{k\in\partial_E i\setminus j}a_{ik}p_{k\to i}\right],
\end{equation}
\begin{equation}\label{activation_nobacktrackE1}
p_{i}=(1-mp_{i})\left[\eta+(1-\eta)\sum_{k\in\partial_E i}a_{ik}p_{k\to i}\right],
\end{equation}
where $1\leq i\leq N_e$, $1\leq j\leq N_e$ and $\partial_E i$ is the set of excitatory neighbors of node $i$. To solve the self-consistent equations, we introduce two auxiliary variables: $G_{i\to j}(\eta,p_{\to})=\eta+(1-\eta)\sum_{k\in\partial_E i\setminus j}a_{ik}p_{k\to i}$, $G_{i}(\eta,p_{\to})=\eta+(1-\eta)\sum_{k\in\partial_E i}a_{ik}p_{k\to i}$. We rearrange Eqs~(\ref{activation_nobacktrackE})-(\ref{activation_nobacktrackE1}) and obtain
\begin{equation}\label{G}
p_{i\to j}=\frac{G_{i\to j}(\eta,p_{\to})}{mG_{i\to j}(\eta,p_{\to})+1}
\end{equation}
and
\begin{equation}
p_{i}=\frac{G_{i}(\eta,p_{\to})}{mG_{i}(\eta,p_{\to})+1}.
\end{equation}
For $\eta=0$ (that is, without external forcing), Eq~(\ref{G}) has a trivial solution: $p_{i\to j}=0$ for all links $i\to j$. The stability of this solution determines the critical state of the system. If the solution is stable, the network activity triggered by a weak stimulus will eventually disappear; otherwise, the response will maintain at a nonzero level.

The stability of the zero solution depends on the Jacobian matrix $\widehat{\mathcal{M}}^E=\{\mathcal{M}^E_{k\to l,i\to j}\}$ defined on all pairs of links $k\to l$ and $i\to j$ between E nodes. Specifically, we have
\begin{eqnarray}
\label{partial_pij}
\mathcal{M}^E_{k\to l,i\to j}&=&\frac{\partial p_{i\to j}}{\partial p_{k\to l}}\bigg\vert_{\{\eta=0, p_{i\to j}=0 \}} \nonumber \\
&=&\frac{\frac{\partial G_{i\to j}}{\partial p_{k\to l}}(mG_{i\to j}+1)-mG_{i\to j}\frac{\partial G_{i\to j}}{\partial p_{k\to l}}}{(mG_{i\to j}+1)^2}\nonumber \\
&=&\frac{\partial G_{i\to j}}{\partial p_{k\to l}}\bigg\vert_{\{\eta=0, p_{i\to j}=0 \}}.
\end{eqnarray}
Here $G_{i\to j}=0$ when $\eta=0$ and $p_{i\to j}=0$ for all $i\to j$. According to the definition of $G_{i\to j}$, the partial derivative of $G_{i\to j}$ is given by
\begin{equation}\label{partial_G}
\frac{\partial G_{i\to j}}{\partial p_{k\to l}}\bigg\vert_{\{\eta=0, p_{i\to j}=0 \}}=\bigg\{
\begin{array}{cc}
a_{lk} &\text{if $l=i$ and $j\neq k$,}\\
0   &\text{otherwise.} \\
\end{array}
\end{equation}
The elements of $\widehat{\mathcal{M}}^E$ are $\mathcal{M}^E_{k\to l,i\to j}=a_{lk}$ if $l=i$ and $j\neq k$ and 0 otherwise. Note that, $\mathcal{M}^E_{k\to l,i\to j}$ is non-zero only if the links $k\to l$ and $i\to j$ are consecutive ($l=i$) but not backtracking ($j\neq k$). The weighted non-backtracking (WNB) matrix, or Hashimoto matrix~\cite{hashimoto1989zeta}, has recently found applications in several problems in network science~\cite{morone2015influence,martin2014localization,karrer2014percolation,hamilton2014tight,teng2016collective,wang2018optimal,wang2019stability,aleja2019nonbacktracking}. Because the stability of the zero solution is determined by the largest eigenvalue $\lambda^E_{NB}$ of $\widehat{\mathcal{M}}^E$, the system reaches the critical state if  $\lambda^E_{NB}=1$.

\subsubsection{Numerical results for dynamics with inhibition}

\begin{figure}
\centering
\includegraphics[width=0.8\columnwidth]{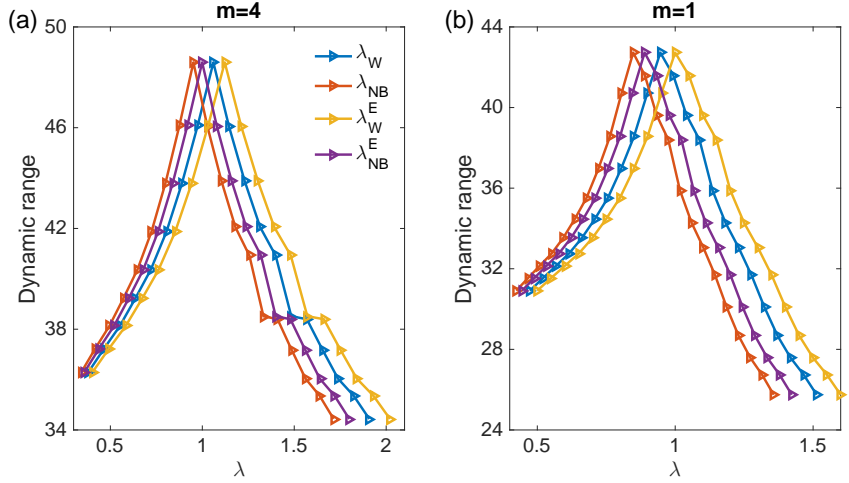}\\
\caption{The relationship between dynamic range and the largest eigenvalues of four matrices for dynamics with 3 refractory states (a) and without refractory state (b). Here, $\lambda_W$, $\lambda_{NB}$, $\lambda^E_{W}$ and $\lambda^E_{NB}$ are the largest eigenvalues of the weighted adjacency matrix for all nodes, the weighted non-backtracking matrix for all nodes, the weighted adjacency matrix for E nodes, and the weighted non-backtracking matrix for E nodes, respectively. We perform the experiment on an EI network constructed using two ER networks ($N_e=3,000$, $N_i=2,000$, $\alpha=3\times 10^{-3}$, $\beta=1\times 10^{-3}$), and vary link weights to change the state of the system. For each setting of link weights, we calculate the dynamic range and the corresponding largest eigenvalues. The setting that maximizes the dynamic range corresponds to the critical state. We use this numerical method to find the critical state of an EI network. At criticality, we find that $\lambda^E_{NB}$ is close to one for $m=4$ and $\lambda^E_{W}$ is close to one for $m=1$. $\lambda_W$ ($\lambda_{NB}$) is always smaller than $\lambda^E_{W}$ ($\lambda^E_{NB}$) due to the existence of inhibitory nodes.}\label{DR_ER}
\end{figure}

For the general case where inhibition cannot be neglected, it is challenging to derive the analytical condition of criticality from  Eq~(\ref{activation_nobacktrack}). As a result, we have to use numerical methods to find the critical state. In particular, we are interested in how the largest eigenvalue of the WNB matrix for E nodes $\lambda^E_{NB}$ changes with the inhibitory strength $\beta$ at the critical state. We treat the increasing level of inhibition as a perturbation to the special case of $\beta=0$, and examine to what extend the critical condition $\lambda^E_{NB}=1$ will remain valid. In order to tune the system to the critical state, for a fixed inhibitory strength $\beta$, we randomly draw absolute link weights $|a_{ij}|$ from a uniform distribution between 0.1 and 0.2, and then multiply $|a_{ij}|$ with a varying constant until the dynamic range of the system is maximized (i.e., the critical state is reached). The largest eigenvalue $\lambda^E_{NB}$ of $\widehat{\mathcal{M}}^E$ is then computed using a power method~\cite{saad2011numerical}. In figure~\ref{DR_ER}, we show the relationship between dynamic range and the largest eigenvalues of four matrices (i.e.,  the weighted adjacency matrix for all nodes, the weighted non-backtracking matrix for all nodes, the weighted adjacency matrix for E nodes, and the weighted non-backtracking matrix for E nodes). The curves present the largest eigenvalues of different matrices at the critical state where dynamic range is maximized.

\begin{figure}
\centering
\includegraphics[width=1\columnwidth]{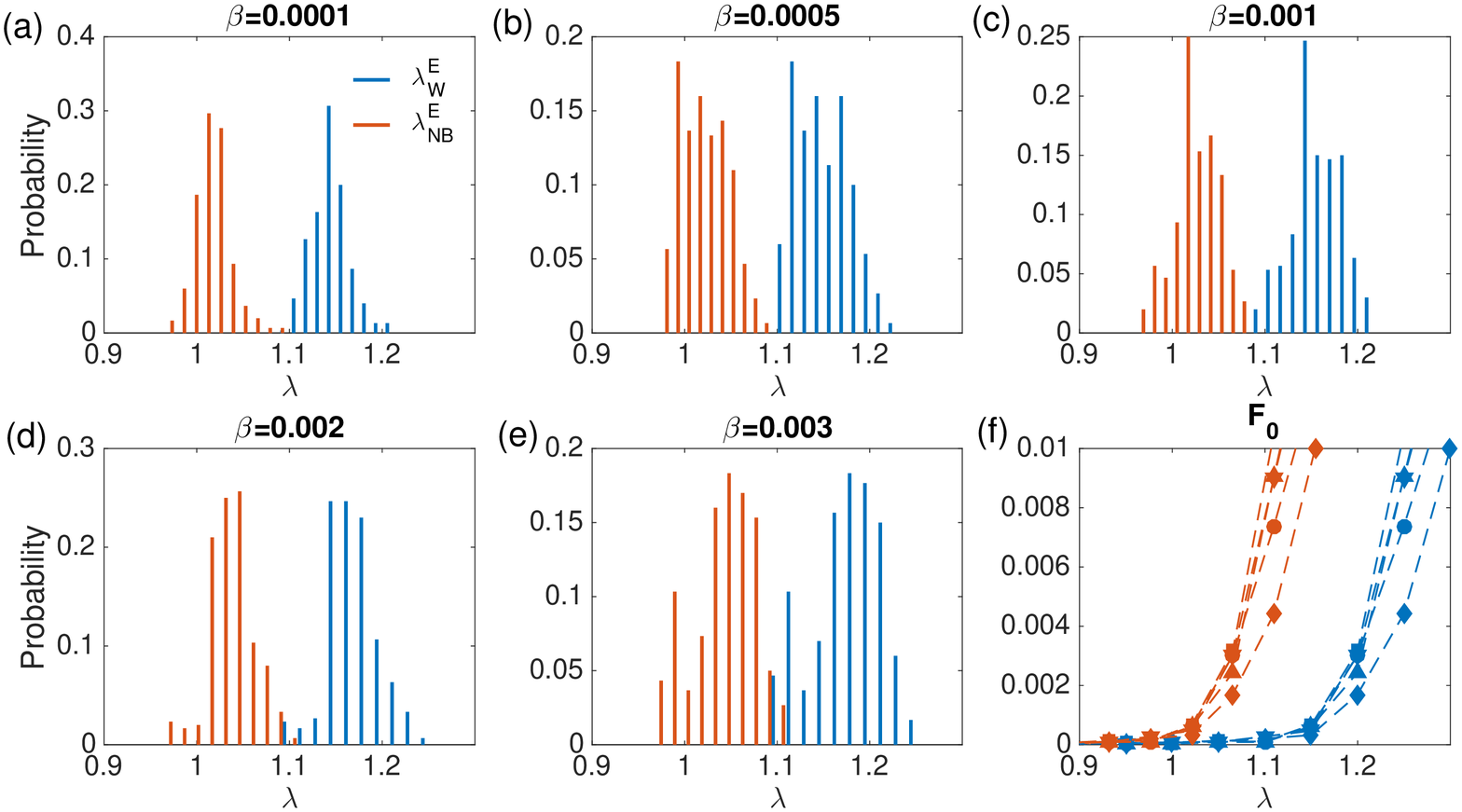}\\
\caption{Distributions of $\lambda^E_W$ and $\lambda^E_{NB}$ at the critical state of homogeneous EI networks with refractory states ($m=4$) (a-e). Networks are constructed by connecting two ER networks ($N_e=3,000$, $N_i=2,000$, $\alpha=3\times 10^{-3}$), and varying the cross-type link probability $\beta$. For different settings of $\beta$, $\lambda^E_{NB}$ is consistently distributed near one. The relationship between $F_0$ and $\lambda^E_W$ and $\lambda^E_{NB}$ is shown in (f) for $\beta=1\times 10^{-4}$ (up triangle), $5\times 10^{-4}$ (square), $1\times 10^{-3}$ (down triangle), $2\times 10^{-3}$ (circle), and $3\times 10^{-3}$ (diamond). The transition point from $F_0=0$ to $F_0>0$ is not affected by the strength of inhibition $\beta$.}\label{ER5}
\end{figure}

We first analyze homogeneous network structure. Without loss of generality, we assume there are 3 refractory states ($m=4$). For ER networks with $N_e=3,000$ excitatory nodes and $N_i=2,000$ inhibitory nodes, we set the within-type connection probability $\alpha=3\times 10^{-3}$. An increasing level of inhibitory strength $\beta=1\times10^{-4}$, $5\times10^{-4}$, $1\times10^{-3}$, $2\times10^{-3}$ and $3\times10^{-3}$ are tested. For each $\beta$, we slowly tune the link weights to drive the system to the critical state, and record the largest eigenvalue of the WNB matrix for E nodes $\lambda^E_{NB}$. We perform 300 realizations of this procedure, and report the distributions of $\lambda^E_{NB}$ in figure~\ref{ER5}. For comparison, we also computed the largest eigenvalue of the weighted adjacency matrix for E nodes $\lambda^E_W$ at criticality.

Interestingly, even with non-negligible inhibition, $\lambda^E_{NB}$ consistently distributes around one at the critical state for an increasing inhibitory strength $\beta$. In contrast, $\lambda^E_W$ distributes well above one. We note that the variation of $\lambda^E_{NB}$ and $\lambda^E_W$ is attributed to the finite network size and numerical inaccuracy, as pointed out in a previous study on excitable networks with only E nodes~\cite{zhang2018dynamic}. The numerical results in figure~\ref{ER5} indicate that the criticality of EI networks with refractory states occurs when $\lambda^E_{NB}$ is close to one, regardless of the strength of inhibition $\beta$. A closer inspection of figure~\ref{ER5} reveals that the average value of $\lambda^E_{NB}$ is slighted larger than one and slowly increases with $\beta$. This slight shift of $\lambda^E_{NB}$ indicates that inhibition does impact the critical condition but its impact is very limited.

According to model dynamics, the function of I nodes is passive -- they need to be first activated before they can release inhibition signals. Without external stimuli, the conduction of inhibitory signals proceeds as follows: a set of excited E nodes activate an I node at time $t$; at time $t+1$, the excited I node exerts inhibitory signals to all its neighbors, among which the excited E nodes enter refractory state. With the presence of nodes in refractory state, we hypothesize that the inhibition effect is weakened. To demonstrate this, we plot $F_0$ as a function of $\lambda^E_{NB}$ and $\lambda^E_W$ for increasing $\beta$ in figure~\ref{ER5}(f). Although the inhibitory strength $\beta$ intensifies, the $F_0$ curve does not change significantly, especially near the transition point from $F_0=0$ to $F_0>0$. This result directly shows that, for dynamics with refractory states, the impact of inhibitory nodes is rather limited near the critical state.

\begin{figure}
\centering
\includegraphics[width=1\columnwidth]{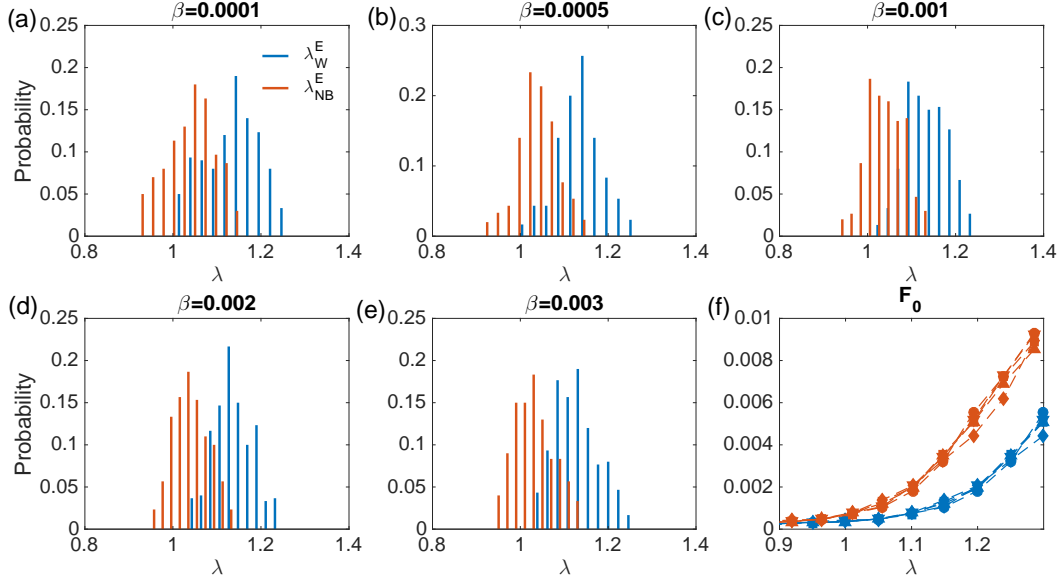}\\
\caption{Distributions of $\lambda^E_W$ and $\lambda^E_{NB}$ at the critical state of heterogeneous EI networks with refractory states ($m=4$) (a-e). Networks are constructed by connecting two SF networks ($N_e=6,000$, $N_i=4,000$, $\gamma=3$), and varying the cross-type link probability $\beta$. For different settings of $\beta$, $\lambda^E_{NB}$ is consistently distributed near one. The relationship between $F_0$ and $\lambda^E_W$ and $\lambda^E_{NB}$ is shown in (f) for $\beta=1\times 10^{-4}$ (up triangle), $5\times 10^{-4}$ (square), $1\times 10^{-3}$ (down triangle), $2\times 10^{-3}$ (circle), and $3\times 10^{-3}$ (diamond). The transition point from $F_0=0$ to $F_0>0$ is not affected by the strength of inhibition $\beta$.}\label{SF5}
\end{figure}

We performed the same analysis in SF networks with $N_e=6,000$, $N_i=4,000$ and the power-law exponent $\gamma=3$. Results in figure~\ref{SF5} show that, consistent with the results for ER networks, $\lambda^E_{NB}$ is close to one at the critical state. The effect of inhibitory nodes near the critical state is also nominal as the $F_0$ curves are almost identical for different values of inhibitory strength $\beta$.

\subsection{Dynamics with backtracking activation}

We now explore the more complicated dynamics in which backtracking activation is allowed. In this case, nodes have only two states -- resting and excited. For each node $i$, denote $p_i^t$ as the probability that node $i$ is excited at time $t$. According to the model dynamics, the evolution of $p_i^t$ is described by
\begin{equation}\label{activation_backtrack}
p_i^{t+1}=(1-p_i^t)\left[\eta+(1-\eta)\sigma\left(\sum_{k=1}^{N}a_{ik}p_{k}^t\right)\right].
\end{equation}
Backtracking activation is properly represented in Eq~(\ref{activation_backtrack}): if we expand $p_k^t$ on the right-hand-side of Eq~(\ref{activation_backtrack}) in terms of the activation probability at time $t-1$, $p_i^{t+1}$ becomes explicitly dependent on $p_i^{t-1}$. This implies, the activation probability of each E node at a given time can contribute to the probability of its re-activation two time-steps later (as long as at least one of its E neighbors are activated), which exactly depicts the effect of backtracking activation.

\subsubsection{Analysis in the case of negligible inhibition}

Similar with our analysis of dynamics with refractory states, we first explore the extreme case where the cross-type linking probability $\beta\to 0$. In this case, we only consider the network of excitatory nodes. The stationary activation probability $p_i=\lim_{t\to\infty}p^t_i$ satisfies
\begin{equation}\label{activation_backtrackE}
p_i=(1-p_i)\left[\eta+(1-\eta)\sum_{k=1}^{N_e}a_{ik}p_{k}\right].
\end{equation}
Without external stimuli, the system has a trivial solution $p_i=0$ for $1\leq i\leq N_e$. The stability of the zero solution is determined by the largest eigenvalue $\lambda^E_W$ of the weighted adjacency matrix for excitatory nodes $A^E=\{a_{ij}\}_{N_e\times N_e}$. As a result, the critical state is characterized by $\lambda^E_W=1$ as $\beta\to 0$.

\subsubsection{Numerical results for dynamics with inhibition}

We perturb the extreme case $\beta\to0$ by gradually increasing the cross-type linking probability $\beta$, which introduces more inhibitory nodes connected to excitatory nodes. Without refractory states, an ``excitatory$\to$inhibitory$\to$excitatory'' feedback loop appears: a group of excited E nodes activate an inhibitory node; the excited I node then releases inhibitory signals and decreases the activation probability of the E nodes who just activated it and now returned to the resting state. The inhibitory signals (negative inputs) impose a threshold for the re-activation of those E nodes. As a consequence, contributions from certain backtracking paths may not be realized. This phenomenon is caused by the threshold-like feature of the transfer function $\sigma(\cdot)$. If the contribution of a backtracking path is lower than the threshold imposed by inhibitory nodes, it may never contribute to the activation probability as $\sigma(x)>0$ only if the net input $x>0$. Following this line of reasoning, Eq~(\ref{activation_backtrackE}) thus overestimates the effect of backtracking activation when more inhibitory nodes are connected to excitatory nodes. A stronger inhibitory strength $\beta$ will suppress more backtracking activations, which drives the dynamics of EI networks closer to the opposite extreme case where backtracking activation is entirely prohibited, described by Eqs~(\ref{activation_nobacktrackE})-(\ref{activation_nobacktrackE1}).

We therefore hypothesize that, for a weak inhibitory strength $\beta$, $\lambda^E_W$ is close to one at the critical state; whereas for a strong inhibitory strength, $\lambda^E_{NB}$ is close to one at the critical state. Varying the cross-type linking probability $\beta$ modulates the system shifting between these two extreme regimes. For an intermediate inhibitory strength $\beta$, we hypothesize that $\lambda^E_{NB}<1$ and $\lambda^E_W>1$ at the critical state. We verify this hypothesis using numerical simulations in both homogeneous and heterogeneous networks.

\begin{figure}
\centering
\includegraphics[width=1\columnwidth]{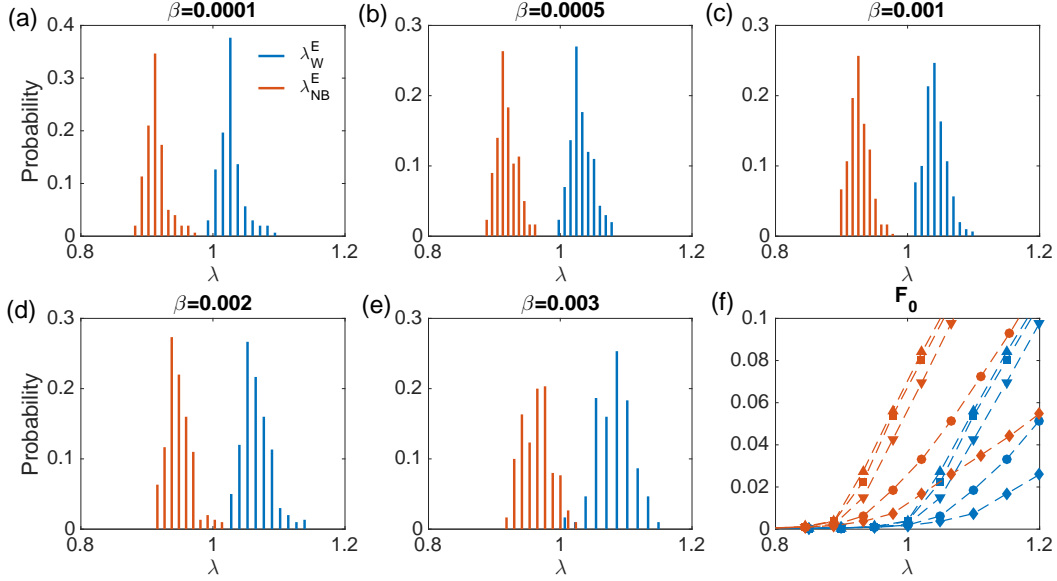}\\
\caption{Distributions of $\lambda^E_W$ and $\lambda^E_{NB}$ at the critical state of homogeneous EI networks without refractory state ($m=1$) (a-e). Networks are constructed by connecting two ER networks ($N_e=3,000$, $N_i=2,000$, $\alpha=3\times 10^{-3}$), and varying the cross-type link probability $\beta$. At the critical state, we show that in general $\lambda^E_W>1$ and $\lambda^E_{NB}<1$. As $\beta$ increases, $\lambda^E_{NB}$ becomes closer to one and $\lambda^E_W$ shifts away from one. The relationship between $F_0$ and $\lambda^E_W$ and $\lambda^E_{NB}$ is shown in (f) for $\beta=1\times 10^{-4}$ (up triangle), $5\times 10^{-4}$ (square), $1\times 10^{-3}$ (down triangle), $2\times 10^{-3}$ (circle), and $3\times 10^{-3}$ (diamond). The transition point from $F_0=0$ to $F_0>0$ is significantly affected by the strength of inhibition $\beta$.}\label{ER2}
\end{figure}

\begin{figure}
\centering
\includegraphics[width=1\columnwidth]{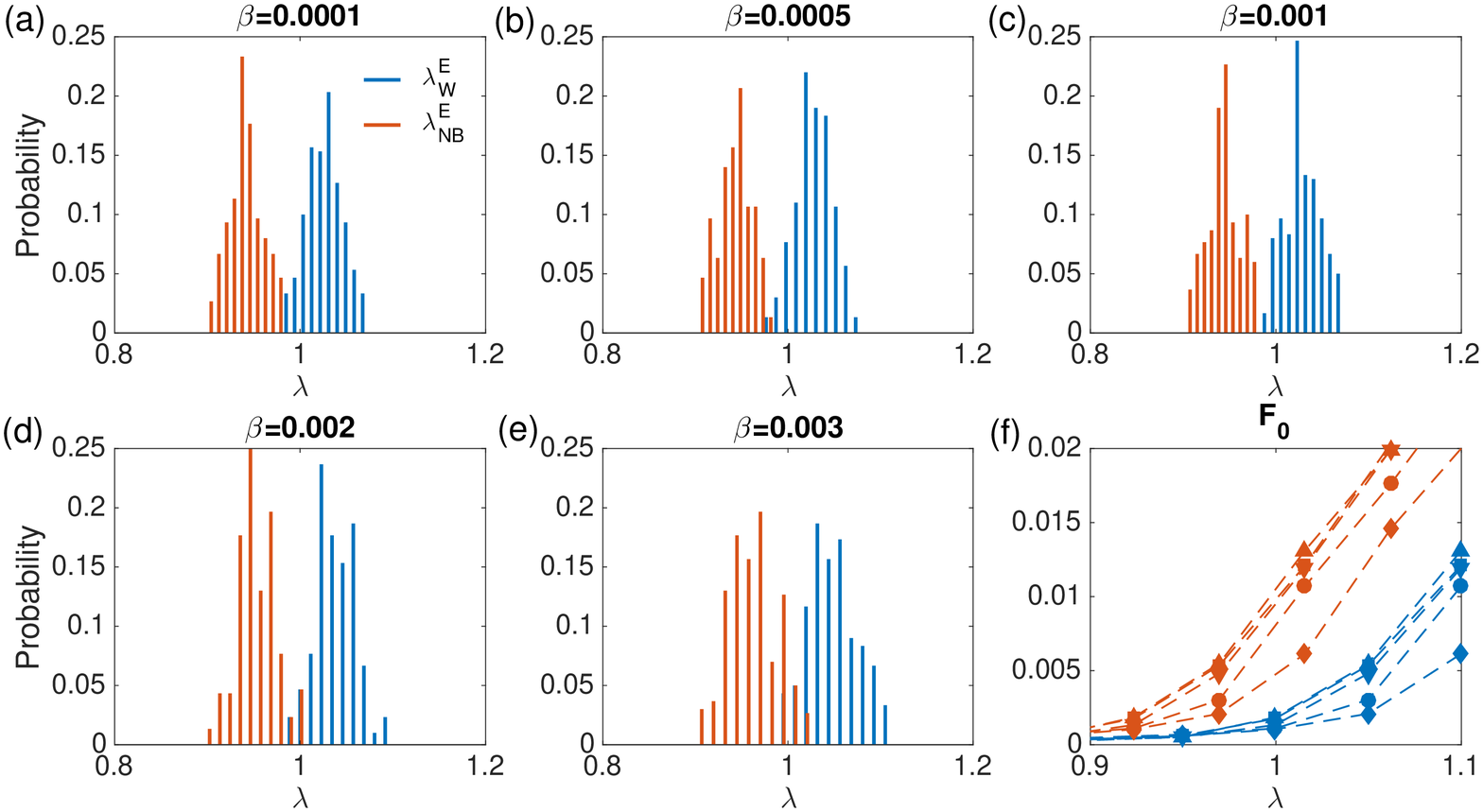}\\
\caption{Distributions of $\lambda^E_W$ and $\lambda^E_{NB}$ at the critical state of heterogeneous EI networks without refractory state ($m=1$) (a-e). Networks are constructed by connecting two SF networks ($N_e=6,000$, $N_i=4,000$, $\gamma=3$), and varying the cross-type link probability $\beta$. At the critical state, we show that in general $\lambda^E_W>1$ and $\lambda^E_{NB}<1$. As $\beta$ increases, $\lambda^E_{NB}$ becomes closer to one and $\lambda^E_W$ shifts away from one. The relationship between $F_0$ and $\lambda^E_W$ and $\lambda^E_{NB}$ is shown in (f) for $\beta=1\times 10^{-4}$ (up triangle), $5\times 10^{-4}$ (square), $1\times 10^{-3}$ (down triangle), $2\times 10^{-3}$ (circle), and $3\times 10^{-3}$ (diamond). The transition point from $F_0=0$ to $F_0>0$ is significantly affected by the strength of inhibition $\beta$.}\label{SF2}
\end{figure}

We performed the same analysis as in figure~\ref{ER5} and figure~\ref{SF5}, except using a different model dynamics with only resting and excited states. The distributions of $\lambda^E_W$ and $\lambda^E_{NB}$ at the critical state for ER networks is shown in figure~\ref{ER2}. In agreement with our hypothesis, as $\beta$ increases, $\lambda^E_W$ shifts from near one to above one, and $\lambda^E_{NB}$ shifts from below one to near one. The same phenomenon is also observed for SF networks in figure~\ref{SF2}. In oder to examine the effect of inhibitory nodes, we plot the $F_0$ curve as a function of $\lambda^E_W$ and $\lambda^E_{NB}$ in figure~\ref{ER2}(f) and figure~\ref{SF2}(f). In contrast to dynamics with refractory states, introduction of more inhibitory links effectively reduces $F_0$, thus strongly impacts the critical state of the system. Such impact is reflected by the change of the transition point above which $F_0$ becomes non-zero.

\begin{figure}
\centering
\includegraphics[width=1\columnwidth]{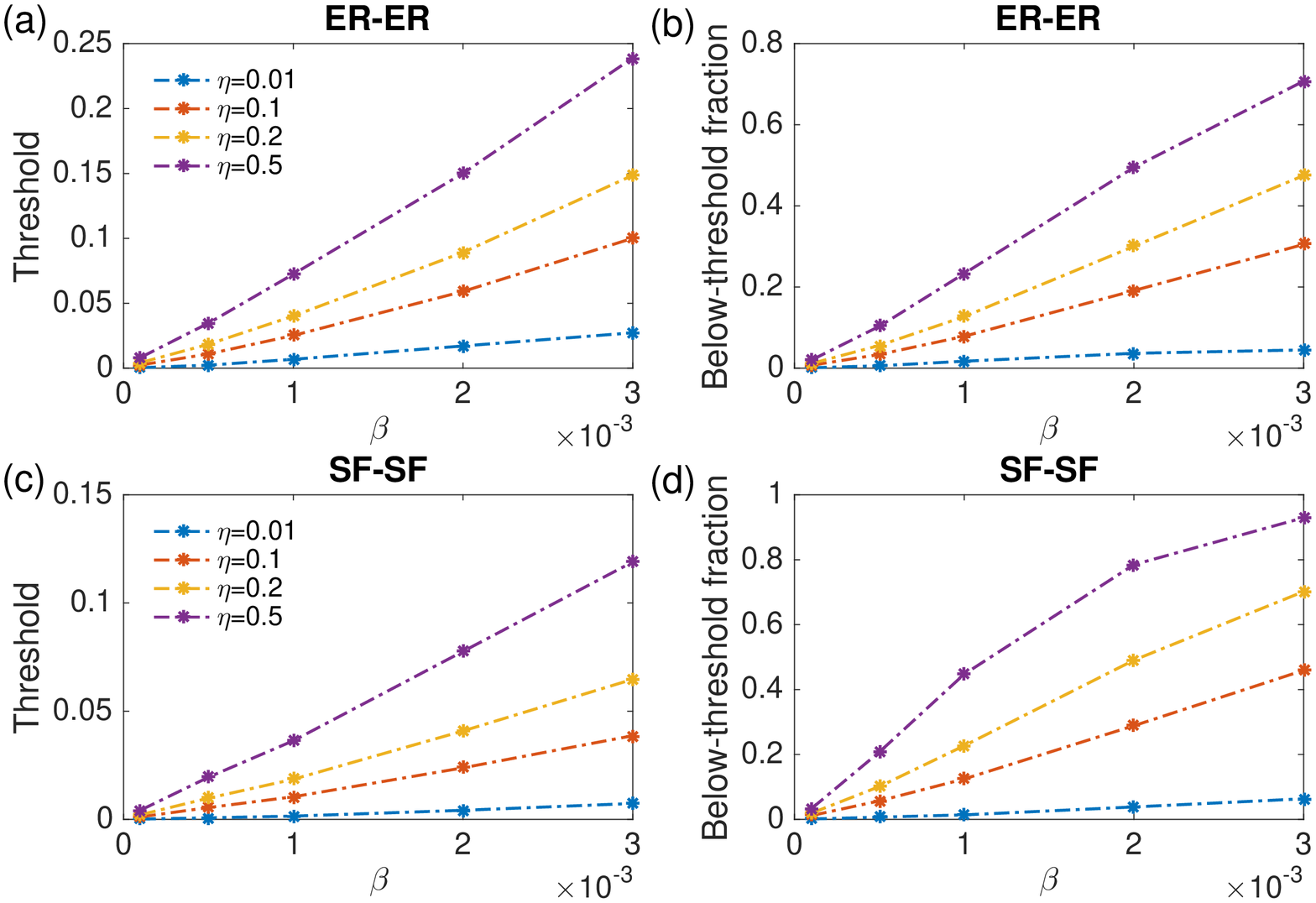}\\
\caption{The threshold of resting E nodes imposed by their inhibitory neighbors for EI networks generated using ER ($N_e=3,000$, $N_i=2,000$, $\alpha=3\times 10^{-3}$) (a) and SF ($N_e=6,000$, $N_i=4,000$, $\gamma=3$) (c) networks. The fraction of below-threshold links for resting E nodes is reported in (b) and (d). For an increasing level of inhibition strength $\beta$, we tune the system to the critical state, and calculate the threshold values and fraction of below-threshold links for different stimulus intensities $\eta$. As $\beta$ and $\eta$ increase, both threshold and below-threshold fraction increase.}\label{threshold}
\end{figure}

Ideally, it would be desirable to show that the number of instances of backtracking activation decreases with an increasing inhibitory strength $\beta$. However, as the activation of a node is collectively determined by a group of nodes, it is difficult to disentangle such interaction and identify definitively which backtracking path is responsible for the activation. Despite that, the impact of inhibitory nodes can be reflected by the threshold values that they impose on excitatory nodes. We calculate the average input from I nodes to E nodes in ER and SF networks. Specifically, for a given stimulus intensity $\eta$, we compute the average input of resting E nodes from their excited inhibitory neighbors at each time step, and then average over all time steps. In figure~\ref{threshold}(a) and (c), we show that the average threshold indeed increases monotonically with $\beta$. In addition, a stronger external stimulus $\eta$ leads to a higher average threshold due to a larger number of excited nodes.

\begin{figure}
\centering
\includegraphics[width=0.6\columnwidth]{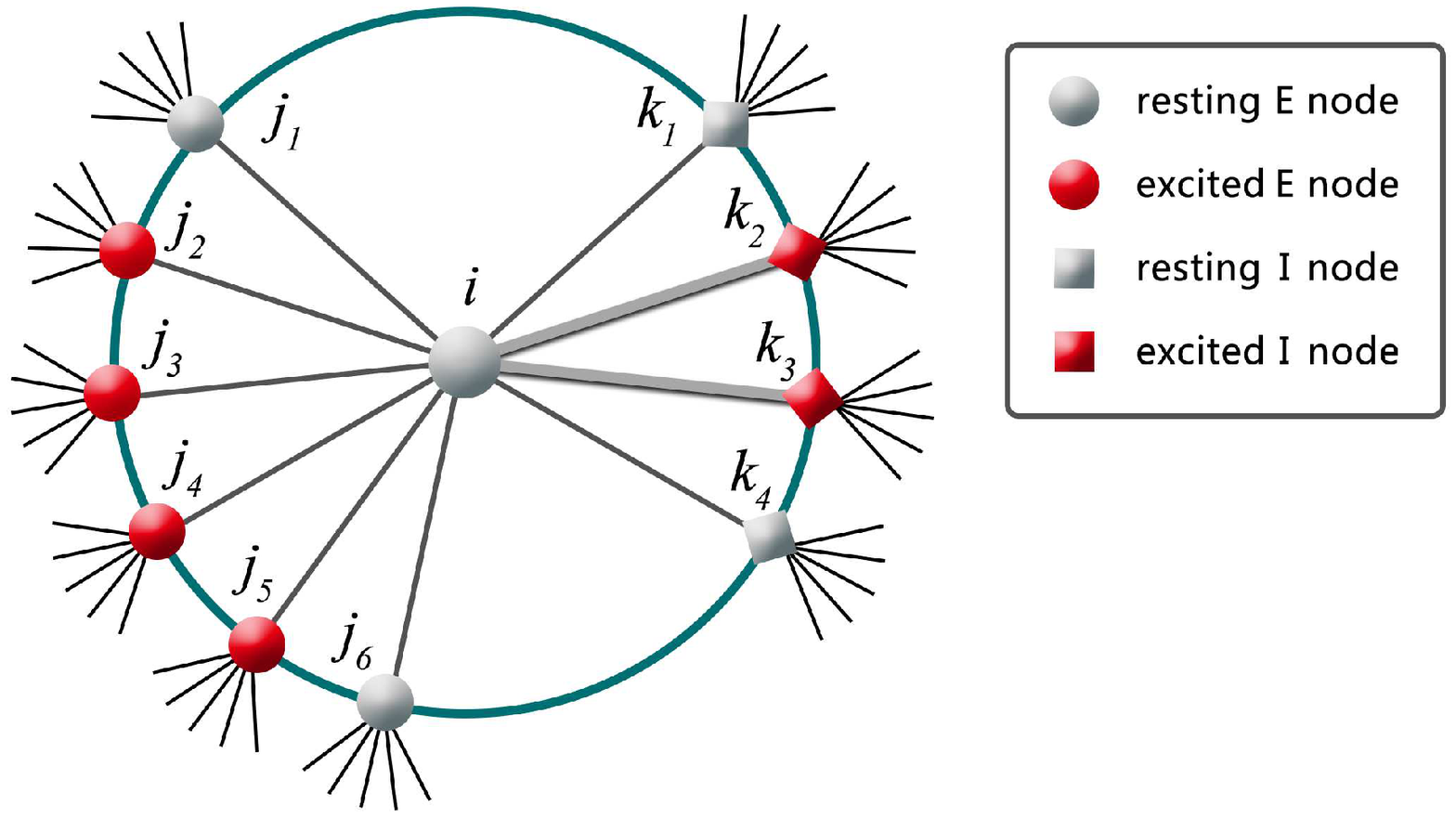}\\
\caption{An example to show the calculation of threshold and fraction of below-threshold links for a resting E node $i$. Here, the resting E node has a total input $|a_{ik_2}+a_{ik_3}|$ for its activated I neighbors. This value is defined as the threshold. Among the $4$ links connected to its activated E neighbors, $2$ links have weights below the threshold ($a_{ij_3}<|a_{ik_2}+a_{ik_3}|$, $a_{ij_4}<|a_{ik_2}+a_{ik_3}|$). The fraction of below-threshold links is calculated as $2/4=0.5$.}\label{example}
\end{figure}

We further calculate the fraction of excitatory links connected to resting E nodes whose weights are lower than the threshold. To be specific, for each resting E node, we find its excited excitatory neighbors and focus on the links connected to them. These links are potential candidates of backtracking activation, i.e., the actual backtracking activation paths belong to this set of links. Among these links, we calculate the proportion whose weights are lower than the threshold of the E node. The contribution from such below-threshold links are likely to be negated by the threshold. Therefore, the fraction of below-threshold links can partly reflect the magnitude of backtracking suppression. We present an illustration for computing this below-threshold fraction in figure~\ref{example}. The mean fraction values averaged over all resting E nodes in all time steps for ER and SF networks are shown in figure~\ref{threshold}(b) and (d). For both network structures, the fraction of below-threshold links increases as $\beta$ grows with the presence of different levels of external stimuli. This analysis agrees with our hypothesis and partially explains the transition between the two extreme cases.

\subsection{Simulations in synthetic neural networks}

\begin{figure}
\centering
\includegraphics[width=1.0\columnwidth]{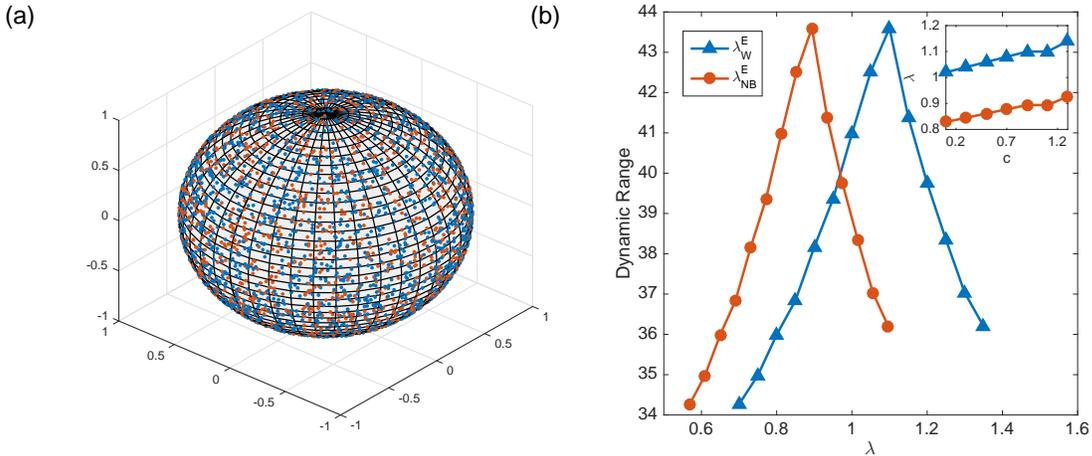}\\
\caption{(a) Generation of the synthetic neural network. 3,000 excitatory nodes (blue) and 2,000 inhibitory nodes (red) are uniformly placed on the surface of a unit sphere. (b) Relationship between the dynamic range and $\lambda^E_W$ and $\lambda^E_{NB}$ for a synthetic neural network ($N_e=3,000$, $N_i=2,000$, $c=1$, the degree distribution follows a normal distribution with a mean of 9 and a variance of 1). We vary link weights to change the state of the system, and calculate the dynamic range, $\lambda^E_W$ and $\lambda^E_{NB}$ for each setting. At the critical state where the dynamic range is maximized, we find $\lambda^E_{NB}<1$ and $\lambda^E_W>1$, which agrees with our hypothesis. Inset shows the values of $\lambda^E_W$ and $\lambda^E_{NB}$ at the critical state for an increasing inhibition strength $c$.}\label{neuralnetwork}
\end{figure}

We further validate our findings in synthetic neural networks that have more realistic structures. As it is difficult to find a real-world neural network dataset that contains both excitatory and inhibitory neurons, we have to construct a synthetic network using network generation models~\cite{stefanescu2008low,markram2004interneurions,wiles2017aupaptic}. In particular, networks of neurons in brain follow a clustered, distance-dependent connection pattern~\cite{wiles2017aupaptic}. We generate networks with this organizational pattern using a distance-dependent method employed in previous studies~\cite{wiles2017aupaptic}. Specifically, $3,000$ excitatory nodes and $2,000$ inhibitory nodes are placed uniformly on the surface of a unit sphere (figure~\ref{neuralnetwork}(a)). The degree of each node was generated from a normal distribution. Nodes were then connected according to a distance-dependent probability $P\propto1/d^2$, where $d$ is the geodesic distance between two nodes on the spherical surface. We assign the synaptic strength of all links from a uniform distribution between 0.1 and 0.2, and multiply a factor $c$ to the weights of cross-type links in order to adjust the strength of inhibition. The value of $c$ reflects the inhibition strength in the system.

We run simulations of model dynamics without refractory state ($m=1$) for $c=1$. We vary link weights, and calculate the dynamic range $\lambda^E_W$ and $\lambda^E_{NB}$ for each weight setting. The relationship between the dynamic range and $\lambda^E_W$ and $\lambda^E_{NB}$ is shown in figure~\ref{neuralnetwork}(b). Similar with the results in random networks, at the critical state, we find that $\lambda^E_W>1$ and $\lambda^E_{NB}<1$. In the inset, we show the values of $\lambda^E_W$ and $\lambda^E_{NB}$ at the critical state for an increasing inhibition strength $c$. As $c$ grows, at the critical state, $\lambda^E_W$ shifts away from one and $\lambda^E_{NB}$ gets closer to one. This result further corroborates our hypothesis that the system lies between two extremes with ($\lambda^E_W\approx 1$) and without ($\lambda^E_{NB}\approx 1$) backtracking activation.

\section{Conclusion}

In this study, we explore the impact of backtracking activation on the criticality of excitable networks with both excitatory and inhibitory nodes. We find that, for dynamics with refractory state that precludes backtracking activation, the critical state occurs when the largest eigenvalue of the WNB matrix for excitatory nodes is close to one. However, for dynamics without refractory state, the introduction of inhibitory nodes affects backtracking activation and the critical condition of the system. The EI model with inhibition essentially provides an intermediate system between two extreme cases in which backtracking activation is allowed or prohibited. For the dynamics with a medium inhibitory strength, $\lambda^E_W$ and $\lambda^E_{NB}$ can be viewed as the upper and lower bound of the critical condition: at the critical state, $\lambda^E_W>1$ and $\lambda^E_{NB}<1$. In practice, this criterion can be used to assess whether a system may be at the critical state. If a system resides in a state where $\lambda^E_W<1$ or $\lambda^E_{NB}>1$, we can assert that the system is not close to the critical state. Our results imply that a precise description of model dynamics is essential in theoretical analysis of phase transitions. These findings highlight the important role of backtracking activation in excitable dynamics.

Critical behavior is common in biological systems~\cite{munoz2018colloquium}. Besides the commonly addressed neuronal networks, fingerprint of criticality have been reported for calcium singallization in myocytes~\cite{nivala2012criticality}, excitable beta cells~\cite{stozer2019heterogeneity}, oocytes~\cite{lopez2012intracellular}, etc. The operation of these biological systems near critical states may be crucial for their proper functioning. Certain inhibitory mechanisms exist in cells to regulate the dynamics of calcium signaling, which could be potentially relevant to our findings in this study. Further, several experimental studies on subcellular~\cite{thul2012subcellular} as well as on the cellular level~\cite{schuster2002modelling} echo our findings that the refractory periods are important in excitable dynamics. Another relevant field is the study on pacemaker activities induced by intracellular calcium waves, which has been found essential in the interstitial cell of Cajal (ICC) in the gastrointestinal tract and cardiac pacemaker cells in the heart. It has been shown that refractory phases are crucial to prevent backtracking of activations in systems guided by pacemaker activities~\cite{gosak2009pacemaker,means2010spatio}. Findings in this study may find applications in these biological systems in future works. It also merits further study whether imposing inhibition on influencers of excitable dynamics would result in efficient regulation of model dynamics~\cite{pei2019influencer,pei2018theories}.

\section*{Data accessibility}
No real-world data were used in this study.

\section*{Authors' contributions}
R.Z. and S.P. designed the study. R.Z. and G.Q. wrote the code, ran simulations and performed the analysis. S.P. wrote the first draft of the manuscript. R.Z., G.Q. and J.W. reviewed and edited the manuscript.

\section*{Competing interests}
We declare we have no competing interests.

\section*{Funding}
This work was supported by the National Natural Science Foundation of China (11801058), the Fundamental Research Funds for the Central Universities (DUT18RC(4)066) and Beijing Natural Science Foundation (1192012, Z180005).

\end{document}